\documentstyle[aaspp4]{article}
\def\p{$\pm$}
\def\l{$\lambda$}
\def\ab{$\sim$}
\def\kms{km s${}^{-1}$}
\def\beam{beam$^{-1}$}

\begin{document}

\title{Radio Detections of Stellar Winds from the Pistol Star and 
Other Stars in the Galactic Center Quintuplet Cluster}
\author{Cornelia C. Lang\altaffilmark{1,2}, Don F. Figer\altaffilmark{2},
W.M. Goss\altaffilmark{1}, and Mark Morris\altaffilmark{2}}
\altaffiltext{1}{National Radio Astronomy Observatory, Box 0, Socorro,
NM 87801, email: clang@nrao.edu}
\altaffiltext{2}{Division of Astronomy, 8371 Math Sciences Building, Box
951562, University of California at Los Angeles, LA, CA 90095-1562}

\begin{center}
To appear in {\it The Astronomical Journal}
\end{center}

\begin{abstract}
VLA images of the Sickle and Pistol H II regions near the Galactic center
at 3.6 and 6 cm reveal six point sources in the region where the
dense Quintuplet stellar cluster is located. 
The spectral indices of five of these sources between 6 cm and 3.6 cm have values of $\alpha$~=~+0.5 to +0.8, (where S$_{\nu}$ $\propto$
$\nu^{\alpha}$), consistent with the interpretation
that the radio sources correspond to ionized stellar winds of
the massive stars in this cluster.  The radio source associated with the 
Pistol Star shows $\alpha$~=~$-$0.4\p0.2, consistent with a flat or slightly
non-thermal spectrum.
\end{abstract}

 
\section{Introduction}
High resolution near-infrared (near-IR) observations over the past decade
have revealed the presence of massive, unusual stars in the inner
50 pc of the Galaxy, where observations suffer 20$-$30
visual magnitudes of obscuration.  Three clusters of
massive stars have been discovered (Glass et al. 1990; Nagata et
al. 1990; Okuda et al. 1990; Krabbe et al. 1995; Nagata
et al. 1995; Figer et al. 1995, Cotera et al. 1996): (1) the Central cluster, located within a parsec of SgrA*; (2) the Arches cluster,
located \ab30 pc N of SgrA* at $\ell$=0\fdg12, b=0\fdg02, and (3) the Quintuplet cluster, also
located \ab35 pc N of SgrA*, at $\ell$=0\fdg16, b=0\fdg06.  The
stars detected in these clusters have near-IR signatures of
OB supergiants and Wolf-Rayet type stars. 
In the densely populated Quintuplet cluster
alone, at least 8 Wolf-Rayet and over a dozen OB
supergiants have been discovered. Based on the evolutionary
stages of the stars, this cluster is likely to be 3.5 Myr old, with a
total estimated mass of \ab10$^4$ M$_{\sun}$, and a mass density of a few
thousand M$_{\sun}$ pc$^{-3}$ (Figer et al. 1999a).   

The near-IR emission line spectra of stars in the three Galactic
center clusters indicate that these stars have evolved away from the zero-age main
sequence and have high-velocity stellar winds with terminal wind speeds of 500$-$1000
\kms~(Nagata et al. 1995, Figer et al. 1999a, Cotera et al. 1996, Krabbe et al. 1995,
Tamblyn et al. 1996). These powerful winds should be detectable at radio
wavelengths, as the radio emission is thermal in nature
(i.e., free-free) and arises from the outer parts of the ionized wind envelope.
The classic theory of Panagia \& Felli (1975) and Wright \& Barlow
(1975) predicts that in the radio regime, the spectrum of wind
emission is proportional to $\nu^{+0.6}$ for a spherically symmetric, isothermal,
stationary wind expanding at a constant velocity.  Previous surveys
made with the Very Large
Array (VLA)\footnotemark\footnotetext{The VLA is a
facility of the National Science Foundation, operated under a
cooperative agreement by the Associated
Universities, Inc.} have detected radio emission arising from the ionized
winds surrounding OB supergiants and Wolf-Rayet stars (Abbott et
al. 1986; Bieging et al. 1989). 

VLA continuum images at 6 cm (4.9 GHz) and 3.6 cm (8.3 GHz) of the Sickle and Pistol H II regions near the Galactic
center reveal six point sources located in the vicinity of the
Quintuplet cluster including the radio source at the position of the
Pistol Star (Lang et
al. 1997; Yusef-Zadeh \& Morris 1987). 
The coincidence of the Pistol star in the near-IR with a peak in the 6
cm radio continuum image was first noted by Figer et al. (1998). 
In this paper we report that two of the newly identified radio sources
in addition to the Pistol Star source are found to be coincident in position
with massive stars in the \l=2.05 $\mu$m {\it HST}/NICMOS image of the Quintuplet cluster
(Figer et al. 1999b). We discuss the nature of the radio sources and
the association with stellar sources in the {\it HST}/NICMOS image.  

\section{Observations}
\subsection{VLA Continuum Observations}
Table 1 summarizes the VLA continuum images in which the radio point sources are detected.
Standard procedures for data reduction and imaging in AIPS have been
used in all cases.
Both images were made with uniform
weighting and have been corrected for primary beam attenuation. The 6 cm continuum image was made
with the data published in Yusef-Zadeh \& Morris (1987), observed with
the VLA in the B, C, and D arrays, and later supplemented with A-array data, to achieve a resolution
of 1\farcs33 $\times$ 1\farcs05, at PA=10\arcdeg. 
  
\subsection{{\it HST}/NICMOS Imaging}
In order to search for stellar counterparts to the radio
point sources, a careful alignment was made between the {\it HST}/NICMOS
$\lambda$=2.05 $\micron$ image of the Quintuplet cluster (Figer et
al. 1999b) and the VLA 8.3 GHz continuum image of this region.  
The Quintuplet cluster was imaged by {\it HST}/NICMOS in a mosaic
pattern in the NIC2 aperture (19\farcs2 on a side)
on UT 1997 September 13/14 in the F205W filter ($\lambda$=2.05 $\micron$).
The MULTIACCUM read mode with NREADS=11 was used for an effective exposure time of 255 seconds per image.
The plate scale was 0\farcs076 pixel$^{-1}$ (x) by 0\farcs075 pixel$^{-1}$ (y), in detector coordinates. 
The cluster was imaged in a 4$\times$4 mosaic, and 
the +y axis of the detector was oriented 135\arcdeg~East of North.
The images were reduced via the standard NICMOS pipeline (CALNICA,
CALNICB; MacKenty et al. 1997) (see Figer et al. 1999b for further details).

A coordinate solution for the {\it HST}/NICMOS image was generated by
assigning known positions to \ab30 Quintuplet cluster stars using the stellar identifications and coordinates from Figer et al. (1999a), which were obtained using the 3-m Shane telescope at
University of California's Lick Observatory.  These near-IR positions
have an absolute accuracy of \ab0\farcs5. The VLA
observations at both 3.6 cm and 6 cm have an uncertainty of only \ab0\farcs1, due
to signal to noise and the known precision of the calibrator
sources. Thus, the alignment of the VLA and {\it HST}/NICMOS images 
has a positional accuracy of $\sigma$=0\farcs5.

\section{Results}
\subsection{Radio Flux Density Measurements}
Figure 1 shows the 3.6 cm continuum image of Lang et al. (1997) in
the vicinity of the Pistol nebula and the Quintuplet
cluster. The six radio point sources discussed in this paper are
labelled QR1$-$QR5 and the Pistol Star. The crosses in the image represent
the positions of {\it HST}/NICMOS sources that are associated with the radio
sources; these associations are further discussed below in $\S$3.2.

In previous radio observations, Yusef-Zadeh \& Morris (1987) identified the
Pistol nebula as the prominent pistol-shaped source at the center of
Figure 1; it has a stellar source near its center of curvature,
the Pistol Star (Figer et al. 1998; and references therein).
The H92$\alpha$ recombination line study of Lang et al. (1997)
characterizes the Pistol nebula as having an electron temperature of T$_e$=3300 K, a complex
velocity structure with central velocity near v$_{LSR}$ \ab120 \kms,
and extremely broad lines ($\Delta$v\ab60 \kms). In addition, a
possible detection of He92$\alpha$ was made, with a helium to hydrogen
abundance, Y$^+$=14\p6\%.  
The continuum emission from the Pistol nebula suggests an H II mass of 11 M$_{\sun}$.   
The absence of molecular material associated with the Pistol nebula, coupled
with the low value of T$_e$ compared to other Galactic center H II
regions, suggest that this nebula may in fact be the ejecta from a
previous stage of the Pistol Star's evolution (Figer et al. 1995; 1998).  
The source of ionization of the Pistol nebula is primarily due to the
radiation field from several of the
Quintuplet cluster members (Figer et al. 1995, 1999c; Lang et al. 1997).

With an rms noise level in the 3.6 cm image of 0.2 mJy \beam, 
the six point sources shown in Figure 1 (QR1$-$QR5 and the Pistol Star
source) are detected with S/N ratios
between 5 and 10. These sources are also detected at 6 cm with S/N ratios
between 5 and 8.  In order to calculate the flux densities at both
frequencies, cross cuts were made in both RA and DEC across each point source. Table 2 lists the 
positions of the point sources, the flux densities at each wavelength,
and the spectral 
index and the deconvolved source size derived from these measurements. 
The radio sources QR1$-$QR5 have rising spectral indices,
$\alpha$~=~+0.5\p0.4 to +0.8\p0.4, (where S$_{\nu}$ $\propto$
$\nu^{\alpha}$), whereas the Pistol Star has a spectral index of
$\alpha$~=~$-$0.4\p0.2, consistent with a flat or slightly
falling spectrum.  

\subsection{{\it HST}/NICMOS Counterparts to VLA sources}

The crosses in Figure 1 show three {\it HST}/NICMOS sources (q15,
q10, and the Pistol star) which are likely associated with the radio
point sources QR4, QR5, and the Pistol Star. The angular offsets in
the radio/near-IR positions are $\lesssim$ 3$\sigma$; the error in the
alignment is dominated by the uncertainty in the near-IR positions of
$\sigma$=0\farcs5. Figure 2 shows the overlay between the {\it HST}/NICMOS
image and the 8.3 GHz continuum image (shown in Figure 1). It is also apparent
in Figure 2 that three of the radio sources (QR4, QR5 and Pistol Star source) are
coincident with {\it HST}/NICMOS sources, and that three of the radio
sources (QR1, QR2 and QR3) do not have {\it HST}/NICMOS counterparts.
Given the relatively large surface density of stars in the {\it
HST}/NICMOS image of the Quintuplet cluster, the possibility of a
chance superposition between a radio source and any {\it HST}/NICMOS
source is not negligible. However, the probability is much smaller
that a randomly placed radio source with a flux density $>$ 1 mJy (5
$\sigma$) is coincident with a near-IR
source that has been classified as a hot, massive star
with a high mass-loss rate (16 sources total in a
77\arcsec~$\times$ 74\arcsec~region of the {\it HST}/NICMOS image;
c.f. Figer et al. 1999b). Excluding the Pistol Star as a special case,
we calculate 
that the combined probability that 2 out of 5 radio sources would be randomly 
aligned (within the 3$\sigma$ positional uncertainty of
1\farcs5) with one of the 16 near-IR supergiants is 4 $\times$ 10$^{-5}$. Therefore, it is highly unlikely 
that these coincidences are due to chance superposition, and indeed 
represent real associations.   
 
\section{Discussion}
\subsection{The Nature of the Radio Sources}

The near-IR counterparts of QR4 and QR5 have been classified as
hot, massive stars with high mass-loss rates: q15 has been classified
as an OB I supergiant and q10 as WN9/Ofpe, according to Figer et al. (1999a).
The radio point sources QR4 and QR5 are presumably detections of the
stellar winds arising from the near-IR stars, since their spectra are
consistent with $\nu$$^{+0.6}$ and they have near-IR counterparts. 
In addition, based on the classic theory of Wright \& Barlow (1975) and Panagia \&
Felli (1975), it is possible to predict the radio flux density of the
stellar wind arising from an OB supergiant found in the Quintuplet cluster
at 3.6 cm.  Assuming the following wind parameters (near the extreme
values) for an OB supergiant at the
Galactic center---a maximum mass loss rate of $\dot{M}$=10$^{-4}$ M$_{\sun}$ yr$^{-1}$, an electron temperature of
T$_e$=25,000 K, a terminal wind velocity of  v$_{\infty}$=500 \kms, and
a distance of 8.0 kpc; the predicted radio flux density at 3.6 cm is \ab4 mJy.  At this frequency, QR1$-$QR5 have flux densities in the range of 2$-$6 mJy,
consistent with this prediction. Given the rms noise in our images of 0.2 mJy
\beam, we are capable of detecting
emission from the winds of OB supergiants in the region of the
Quintuplet cluster, and the radio sources are most likely detections
of these ionized winds.  Since there are at least 8 Wolf-Rayet type stars in the Quintuplet cluster, we
can also estimate the radio flux density at 3.6 cm for these stars, using
the following wind parameters: \.{M}=5 $\times$
10$^{-5}$ M$_{\sun}$ yr$^{-1}$, T$_e$=40,000 K, v$_\infty$=2000 \kms,
and d=8.0 kpc; the predicted flux density for a Wolf-Rayet
star at the Galactic center is \ab0.05 mJy at 3.6 cm.
Since the rms noise in both of the VLA continuum maps is 0.2 mJy
\beam, we would clearly not have detected the mass-losing Wolf-Rayet
stars in the current data, and are only sensitive to the winds arising
from OB supergiants.   

Although the
sources QR1, QR2, and QR3 are detected with S/N $>$ 5, and have
spectral indices consistent with stellar wind sources, they have no
obvious {\it HST}/NICMOS stellar counterparts. A possible
explanation is that the near-IR extinction varies across the cluster,
and that the stellar counterparts of QR1, QR2 and QR3 are masked by
greater extinction than the stellar counterparts of QR4 and QR5.
If we invoke extinction to explain the lack of counterparts for QR1$-$QR3,
then a near-IR extinction A$_k$ $>$ 8 is required, corresponding to
a visual extinction A$_v$ $>$ 80. This kind of extinction is only possible if a dense
molecular cloud is located in front of part of the Quintuplet cluster.
In that case, the unseen counterparts could still be members of the
cluster. However, there is no evidence for such a molecular cloud in
this region, which makes this suggestion unlikely.

\subsection{The Pistol Star}

The spectral index of the Pistol Star ($\alpha$~=~$-$0.4\p0.2) is
consistent with a flat or slightly falling spectrum. It does not follow the
classic theory for a fully ionized wind, which predicts a rising
spectrum, $\alpha$~=~+0.6. The Pistol Star, a prominent
source in the near-IR {\it HST}/NICMOS image, has been
classified as a Luminous Blue Variable (LBV) by Figer et al. (1998)
and has a stellar wind.  Based on the stellar parameters for the
Pistol Star (c.f., Figer et al. 1998, the ``L'' model), 
\.{M}=3.8 $\times$ 10$^{-5}$ M$_{\sun}$ yr$^{-1}$, T$_e$=12,000 K,
v$_{\infty}$=100 \kms, and a distance of 8.0 kpc, the predicted radio
flux density at 3.6 cm is \ab9 mJy using the formulation of Panagia
\& Felli (1973) and Wright \& Barlow (1975).  At 3.6 cm, the flux
density of the Pistol Star is 5.8\p1.0 mJy, and at 6 cm the flux
density is 7.4\p1.0 mJy.

The radio emission of the Pistol Star source is likely a detection of the ionized wind
arising from the Pistol Star. One possible explanation for the
slightly falling spectrum is that the Pistol Star may have a non-thermal
component in its wind over the cm-wavelength range.  This type of
spectral index has been observed from other supermassive stars, with
$\alpha$ in the range of $\alpha$=$-$0.8 to 0.0 (Abbott et al. 1984;
Persi et al. 1985). In fact, the VLA survey of of Galactic OB stars
made by Bieging et al. (1989) finds that 24\% of luminous supergiants
are observed to have non-thermal spectra.  This fraction is consistent
with our results: 1 of the 6 radio point sources we detect has a
slightly falling spectral index. Non-thermal emission is thought to arise either by means
of shocks in the wind itself, in the shock between the stellar wind
and a binary companion (Contreras et al. 1996), or from the
interaction of the stellar wind with the remnant of a star's previous
evolutionary mass-loss phase (Leitherer et al. 1997). 
  
\section{Conclusions} 

Six point sources were detected at 3.6 cm and 6 cm with the VLA in the
vicinity of the Quintuplet cluster. These
sources have rising spectra in the range of $\alpha$=+0.5 to +0.8,
with the exception of the Pistol Star ($\alpha$=$-$0.4).  Based on
the overlay of the {\it HST}/NICMOS and 8.3 GHz VLA continuum images, three of these radio sources, including the Pistol
Star source, can be identified with hot, massive stars with high-mass
loss rates. Therefore, the radio sources are most
likely detections of the ionized stellar winds emanating from the
supermassive stars in the Quintuplet cluster.

We would like to thank Luis Rodriguez for suggesting that the stellar
wind of the Pistol Star may be detectable at 8.3 GHz. We would
also like to thank Liese van Zee for help with the coordinate
solution for the {\it HST}/NICMOS image, and Paco Najarro for useful
comments on the theory of stellar winds.

\clearpage
{\scriptsize
\begin{deluxetable}{lcccc}
\singlespace
\tablecaption{VLA Continuum Image Parameters}
 
\tablehead{\colhead{$\lambda$} &
\colhead{Resolution} &
\colhead{rms noise} &
\colhead{Arrays} &
\colhead{Reference}\\}
\tablecolumns{5}
\startdata
3.6 cm&2\farcs04 $\times$ 1\farcs71, PA=59\arcdeg&0.2 mJy beam$^{-1}$& DnC, Cn
B&Lang, Goss, \& Wood (1997)\\
6 cm&1\farcs33 $\times$ 1\farcs05, PA=10\arcdeg&0.2 mJy
beam$^{-1}$&A,~B,~C,~D&Yusef-Zadeh \& Morris (1987)\\
\enddata
\end{deluxetable}}

{\scriptsize
\begin{deluxetable}{lccccccc}
\tablecaption{Parameters of the Radio Point Sources} 
\tablehead{\colhead{Source} &
\colhead{R.A. (B1950)} &
\colhead{Decl. (B1950)} &
\multicolumn{2}{c}{Flux Density (mJy)} &
\colhead{Spectral}&
\colhead{Deconvolved}&\colhead{NICMOS}\\
\cline{4-5}
\colhead{Name}&
\colhead{(h m s)}&
\colhead{(\arcdeg~\arcmin~\arcsec)}&
\colhead{$\lambda$=3.6 cm} & 
\colhead{$\lambda$=6 cm} &
\colhead{Index\tablenotemark{*}}&\colhead{Size}&\colhead{Counterpart}\\}
\tablecolumns{8}
\startdata
Pistol Star&17 43 04.77\p0.01&-28 48 57.0\p0.1&5.8\p1.0&7.4\p1.0&$-$0.4\p0.2&3\farcs5&yes\\
QR1&17 43 04.72\p0.01&-28 48 14.9\p0.1 &3.0\p1.0&1.9\p0.4&+0.8\p0.3&2\farcs0&no\\
QR2&17 43 04.52\p0.01&-28 48 13.2\p0.1&5.6\p1.0&3.6\p0.9&+0.8\p0.2&2\farcs5&no\\
QR3&17 43 03.95\p0.01&-28 48 12.1\p0.1&3.0\p1.0&2.2\p0.7&+0.8\p0.4&2\farcs0&no\\
QR4&17 43 04.61\p0.01&-28 48 22.2\p0.1&2.9\p0.9&1.8\p0.6&+0.5\p0.4&3\farcs0&yes\\
QR5&17 43 04.60\p0.01&-28 48 30.0\p0.1&1.3\p0.7&0.8\p0.5&+0.8\p0.6&4\farcs0&yes\\
\enddata
\tablenotetext{*}{S $\propto$ $\nu^{\alpha}$}
\end{deluxetable}}

\clearpage
{\centering \leavevmode
\epsfxsize=0.70\columnwidth
\epsfbox{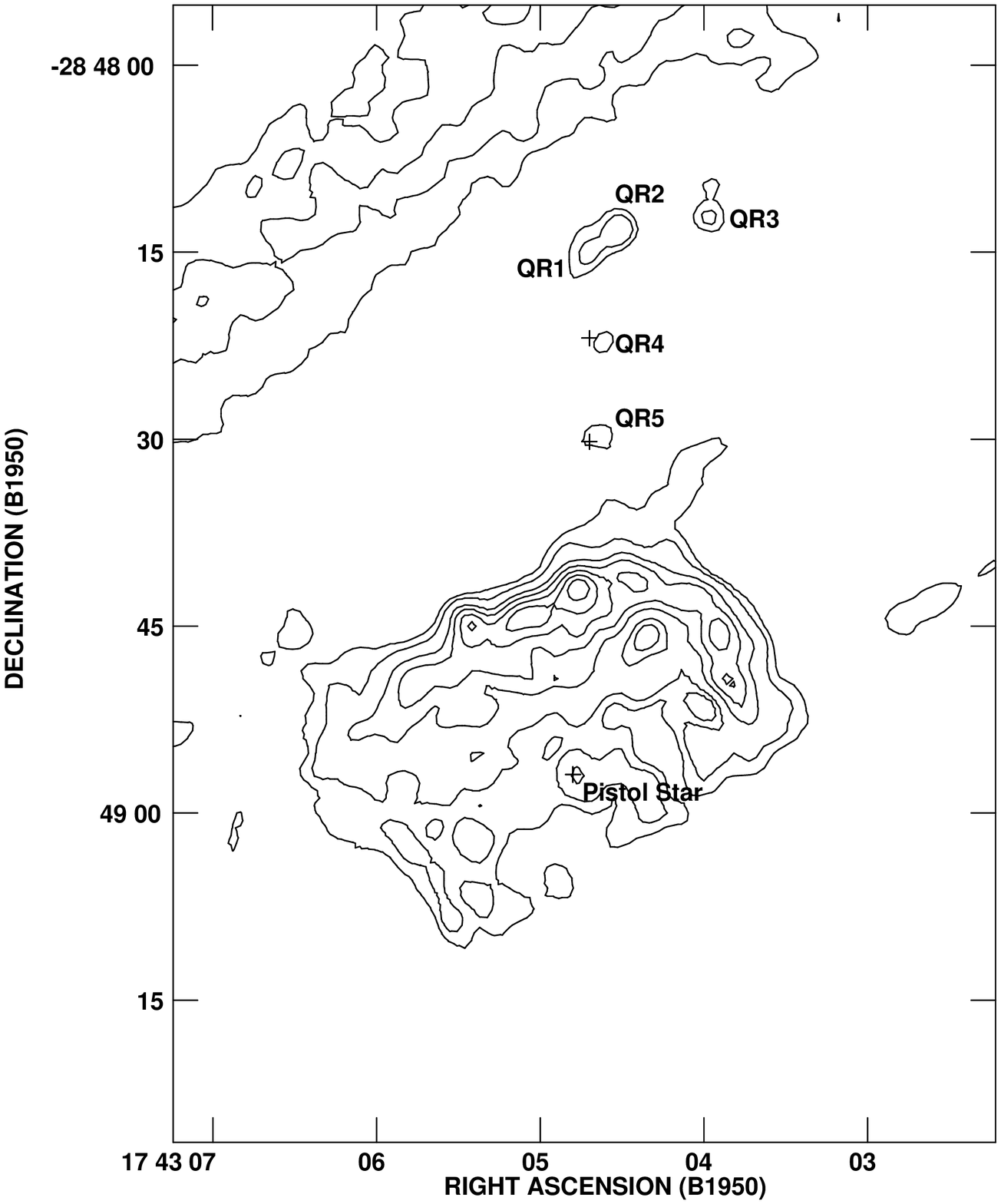}
 
\parbox[b]{0.70\columnwidth}
 
{Figure 1 - VLA 3.6 cm  radio continuum image of the six radio point
sources detected at both 3.6 and 6 cm, labelled QR1$-$QR5, and the
Pistol Star. This image has a resolution of 2\farcs12 $\times$
1\farcs71, PA=59\arcdeg.  The contours represent 0.5, 1, 2.5, 4, 6, 8, 10, 12,
14, 16, 18, 20, 22 mJy $beam^{-1}$; where 0.5 mJy beam$^{-1}$
corresponds to 1.5$\sigma$. The crosses represent {\it HST}/NICMOS
positions of stars that are associated with three of the radio sources, with
positional uncertainties of $\lesssim$ 3$\sigma$, where $\sigma$=0\farcs5.}}

{\centering \leavevmode
\epsfxsize=0.80\columnwidth
\epsfbox{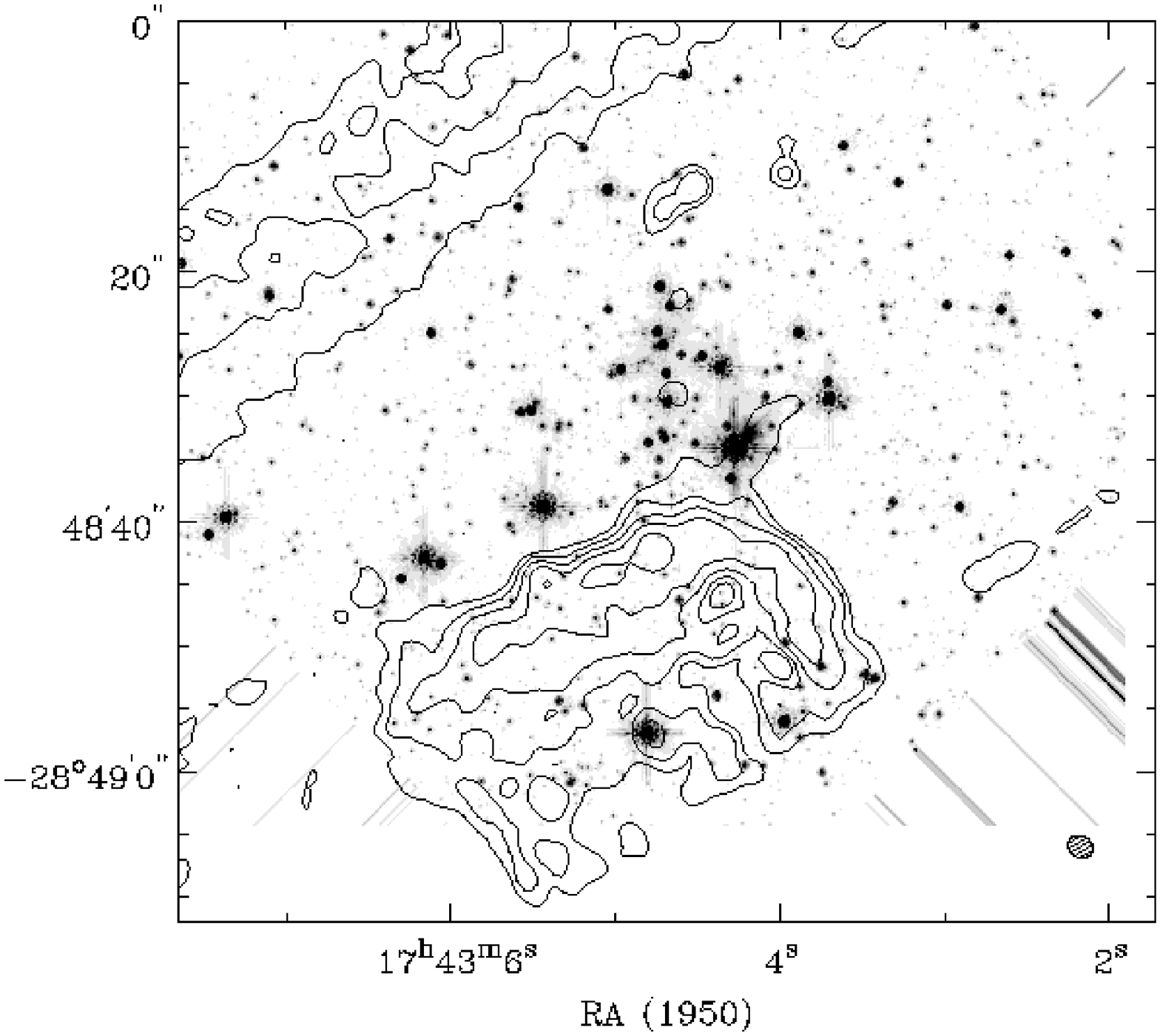}
 
\parbox[b]{0.80\columnwidth}
 
{Figure 2 - VLA 3.6 cm continuum image (as shown in Figure 1)
overlaid on {\it HST}/NICMOS $\lambda$=2.05$\micron$ image of the
Quintuplet cluster.}}

\end{document}